\documentstyle[12pt,a4]{article}

\begin{document}

\baselineskip 24pt

\thispagestyle{empty}

\begin{flushright}
BARI - TH 110/92 \\
September, 1992  \\
\end{flushright}

\vspace{1cm}

\begin{center}
{\large \bf
Dual Superconductor Mechanism of Confinement\\ on the Lattice}
\end{center}

\vspace{0.5cm}

\begin{center}

{P. Cea$^{a,b}$}  and L. Cosmai$^b$ \\[0.2cm]
{\it $^{\rm a}$Dipartimento di Fisica dell'Universit\`a di Bari,
 70126 Bari, Italy\\
{\rm and}\\
$^{\rm b}$Istituto Nazionale di Fisica Nucleare,
Sezione di Bari,70126 Bari, Italy}\\
cea, cosmai  @bari.infn.it

\end{center}

\vspace{0.5cm}

\begin{abstract}

We investigate the dual superconductor mechanism of
confinement for pure SU(2) lattice gauge theory in the
maximally abelian gauge. We focus on the the
dual Meissner effect. We find that the transverse distribution
of the longitudinal chromoelectric field due to a static
quark-antiquark pair satisfies the dual London equation.
Moreover we show that the size of the flux tube scales
according to asymptotic freedom.

\end{abstract}

\vspace{\fill}
\begin{flushright}
HEP-LAT Preprint: hep-lat/9210030
\end{flushright}

\newpage

Long time ago  G.~'t~Hooft and S.~Mandelstam\cite{thooft75} proposed
that the confining vacuum is a coherent state of color magnetic
monopoles. This proposal offers a picture of confinement whose
physics can be clearly extracted. Indeed the dual Meissner effect
causes the formation of chromoelectric flux tubes between
chromoelectric charges leading to a linear confining
potential. Following Ref.~\cite{thooft81}, one can study the monopole
condensation by means of the so called Abelian projection.
It turns out that~\cite{kronfeld87} it is possible to implement
the Abelian projection on the lattice, where the Abelian projection
amounts to fix the gauge by diagonalizing an operator which
transforms according to the adjoint representation of the gauge group.
In this paper we consider the SU(2) gauge group with standard Wilson
action on a $12^4$ lattice in the maximally Abelian gauge~\cite{laursen87}.
In this gauge one diagonalize the lattice operator
\begin{equation}
\label{X(x)}
X(x) =
\sum_{\mu}
\left\{ U_\mu(x) \sigma_3 U_\mu^\dagger(x) +
U_\mu^\dagger(x-\hat\mu) \sigma_3 U_\mu(x-\hat\mu) \right\} \; .
\end{equation}
The gauge is fixed iteratively via overrelaxation like the
Landau gauge~\cite{mandula}. We adopted a convergence criterion
which coincides with the one of Ref.~\cite{hioki91}

The aim of this paper is to analyze the dual Meissner effect
by studying the color fields distribution due to a static
quark-antiquark pair. To measure the color fields we follow
the method of Ref.~\cite{digiacomo}. These authors measure
the correlation of a plaquette $U_P$ with a Wilson loop $W$.
The plaquette is connected to the Wilson loop by a Schwinger
line $L$:
\begin{equation}
\label{rhow}
\rho_W = \frac{ \left\langle \mathrm{tr}
\left( W L U_P L^{\dagger} \right) \right\rangle }
              { \left\langle \mathrm{tr}
\left( W \right) \right\rangle }
 - \frac{1}{2} \,
\frac{ \left\langle \mathrm{tr} \left( U_P \right)
 \mathrm{tr} \left( W \right)  \right\rangle }
              { \left\langle \mathrm{tr}
\left( W \right) \right\rangle } \;\; .
\end{equation}
By moving the plaquette $U_P$ with respect to the Wilson loop one
can scan the structure of the color fields.
Note that in the na\"{\i}ve continuum limit, the operator
$\rho_W$ is  sensitive to the fields rather than
to the square of the fields~\cite{digiacomo}.

Following the Abelian dominance idea~\cite{hioki191} we investigate the
color field distribution by measuring
\begin{equation}
\label{rhowab}
\rho_W^{ab} = \frac{ \left\langle \mathrm{tr}
\left( W^A U_P^A \right) \right\rangle }
              { \left\langle \mathrm{tr}
\left( W^A \right) \right\rangle }
 - \frac{1}{2} \,
\frac{ \left\langle \mathrm{tr} \left( U_P^A \right)
 \mathrm{tr} \left( W^A \right)  \right\rangle }
              { \left\langle \mathrm{tr}
\left( W^A \right) \right\rangle } \;\; .
\end{equation}
where the superscript means that the Wilson loop and the plaquette are
built from the abelian projected links.

We use Wilson loop of size $5$.
We average over $500$ configurations (each one separated by $50$ upgrades,
after $3000$ sweeps to equilbrate the lattice) in the range
$ 2.4 \le \beta \le 2.525 $.

The authors of Ref.~\cite{digiacomo} found a sizeable signal for the
chromoelectric field parallel to the flux tube ($U_P$ parallel to $W$).
So we focus on the longitudinal chromoelectric field. By moving the
plaquette outside the plane of Wilson loop up to distance $5$ in lattice
units, we measure the transverse profile of the longitudinal chromoelectric
field in the middle of the flux tube. In Fig.~1 we show the result for
two different value of $\beta$.

The authors of Ref.~\cite{digiacomo} found that the transverse shape of
the flux tube can be fitted in accordance with
\begin{equation}
\label{gauss}
E_l(x_\perp) = A_G \exp \left[ -m_G x_\perp - \mu_G^2 x_\perp^2 \right] \;,
\;\; x_\perp \ge 0 \;.
\end{equation}
Equation~(\ref{gauss}) describes the flux tube like a relativistic
string with gaussian fluctuations~\cite{lusher81}.

We find that also our data can be fitted by Eq.~(\ref{gauss}). In Fig.~1
the dashed line is the result of the fit~(\ref{gauss})
($A_G$ is fixed by $E_l(x_\perp=0)$). Moreover in Fig.~2 we check
the scaling of $\mu_G$. We find that $\mu_G$ scales and
\begin{equation}
\label{mug}
\frac{\mu_G}{\Lambda_L} = 83 \pm 2 \;\;\;.
\end{equation}
Such nice scaling property is not shared by $A_G$ and $m_G$. The
value~(\ref{mug})
 is quite close to the one $\frac{\mu_G}{\Lambda_L}=75\pm 2$
obtained in Ref.~\cite{digiacomo} by using non abelian quantities on
a $16^4$ cooled lattice. Thus, we feel that this result support the
Abelian dominance idea.

On the other hand, we find that the data are compatible with another
functional form, namely
\begin{equation}
\label{k0}
E_l(x_\perp) = A_M K_0 \left( \mu_M x_\perp \right) \;,
\;\; x_\perp > 0 \;,
\end{equation}
where $K_0$ is the modified Bessel function of order zero. Equation~(\ref{k0})
 is a straigthforward consequence of the dual superconductor hypothesis
(see also the infrared effective theory of the monopole condensation
proposed in Ref.~\cite{suzuki88}).
Indeed, let us consider a second kind superconductor in an external
static magnetic field. If we have an isolated vortex line, then in the London
limit the magnetic field satisfies the London equation~\cite{degennes89}

\begin{equation}
\label{london}
h - \lambda^2 \, \nabla^2 \, h =
\varphi_0 \, \delta^{(2)}(x_\perp) ,
\end{equation}
where $h$ is the magnetic field parallel to the vortex line, $x_\perp$
the transverse distance from the vortex line, and $\varphi_0$  the
magnetic flux. The penetration depth $\lambda$  is related to
the photon mass by the well known relation
\begin{equation}
\label{photonmass}
\lambda = \frac{1}{m_\gamma}  \; .
\end{equation}
The solution of Eq.~(\ref{london}) is
\begin{equation}
\label{lonk0}
h(x_\perp) =  \frac{\varphi_0}{2 \pi} \, \frac{1}{ \lambda^2} \,
K_0\left(\frac{x_\perp}{\lambda}\right) \;\;\;\; x_\perp > 0 \;.
\end{equation}
Interchanging magnetic with electric, we are lead to consider the fit
Eq.~(\ref{k0}) with $\mu_M = \frac{1}{\lambda}$ and
$A_M \sim \mu_M^2$.

Recently, it has been shoved that~\cite{haymaker92} that the longitudinal
electric field of the $U(1)$ flux tube satisfies the dual London equation.
We will compare our method with the one of Ref.~\cite{haymaker92} in a
separate paper.

The results of the fit Eq.~(\ref{k0}) are plotted as solid line in Fig.~1.
Both fits Eqs.~(\ref{gauss}) and ~(\ref{k0}) give a comparable reduced
 $\chi^2$. Moreover $\mu_M$ scales (see Fig.~3):
\begin{equation}
\label{mum}
\frac{\mu_M}{\Lambda_L} = 132 \pm 2 \; .
\end{equation}
It turns out that (see Fig.~4):
\begin{equation}
\label{am}
A_M = (0.23705 \pm 0.00841) \; \mu_M^2 \; .
\end{equation}
Following Ref.~\cite{lusher81} we can define the width of the flux tube by
\begin{equation}
\label{width}
D = \frac{\int \mathrm{d}^2x_\perp\;x_\perp^2\,E_l(x_\perp) }
     {\int \mathrm{d}^2x_\perp\;E_l(x_\perp) } \; .
\end{equation}
{}From Equations~(\ref{width}) and ~(\ref{k0})we obtain
\begin{equation}
\label{d}
D = \frac{2}{\mu_M} \; .
\end{equation}
Using~\cite{digiacomo} $ \Lambda_L = 6.8 \pm 0.2 \mathrm{MeV} $
and Eq.~(\ref{mum})
we get
\begin{equation}
\label{dph}
D = 0.44 \pm 0.02 \; \mathrm{fm}
\end{equation}
which is close to the value estimated in Ref.~\cite{digiacomo}.

In a previous paper~\cite{cea91} we propose a method to measure the abelian
photon mass by means of the connected correlation function of an operator
with the quantum number of the photon. In order to check
Eq.~(\ref{photonmass}) we
recorded also the abelian photon mass. In Figure~(5) we display
$\mu_M/m_\gamma$ versus $\beta$. Even though there are large
statistical fluctuations, mainly due to the abelian photon mass, we see
that the relation Eq.~(\ref{photonmass}) is consistent with Monte Carlo
data.

In conclusion we have showed that the transverse distribution of the
longitudinal chromoelectric field due to a static quark-antiquark pair
satisfies the dual London equation. However, care must be taken of
self-energy effects from the Wilson line. So our results should be
checked on larger lattices. We showed that the London
penetration length scales according to asymptotic freedom. This suggests that
the penetration length is a gauge invariant physical quantity.
Some preliminary results on a larger lattice~\cite{cea92} support
this conclusion. Needless to say, this matter will be deepen
in future studies.

\vspace{0.5cm}

\begin{center}
{\bf Aknowledgements}
\end{center}

\vspace{0.2cm}
We thank A. Di Giacomo, R.W. Haymaker and T. Suzuki for useful discussions.
\newpage

\begin{center}
{\large \bf FIGURE CAPTIONS}
\end{center}
\vspace{1cm}
\newcounter{bean}
\begin{list}
{ {\bf Fig.~\arabic{bean} }  }{\usecounter{bean}
\setlength{\rightmargin}{\leftmargin}
\setlength{\labelwidth=2.5cm} }
\item Transverse distribution of the longitudinal chromoelectric
field at a) $\beta=2.4$ and b) $\beta=2.5$. Dashed and solid lines
refer to Eqs.~(4) and (6) respectively.
\item Asymptotic scaling of $\mu_G$. The dashed line is the fitted
value Eq.~(5).
\item Asymptotic scaling of $\mu_M$. The dashed line is the fitted
value Eq.~(10).
\item The ratio $A_M/m_M^2$ versus $\beta$.
\item The ratio $\mu_M/m_\gamma$ versus $\beta$. The solid
line corresponds to  $\mu_M/m_\gamma=1$.
\end{list}
\newpage

\end{document}